\documentstyle[aps,pre,citesort,epsf]{revtex}

\tightenlines

\begin{document}

\draft

\title{Pattern Formation Near Onset of a Convecting Fluid In an Annulus}

\author{
Berk Sensoy and
Henry Greenside\cite{email-correspondence}\cite{CNCS-address}
}

\address{
Department of Physics\\
Duke University, Durham, NC 27708-0305
}

\date{January 25, 2001}

\maketitle

\begin{abstract}
  Numerical simulations of the time-dependent
  Swift-Hohenberg equation are used to test predictions
  of Cross [Phys.\ Rev.~A {\bf 25}:1065-1076 (1982)]
  that Rayleigh-B\'enard convection in the form of
  straight rolls or of an array of dislocations may be
  observed in an annular domain depending on the values
  of inner radius~$r_1$, outer radius~$r_2$, reduced
  Rayleigh number~$\epsilon$, and the initial state.
  As~$r_1$ is decreased for a fixed~$r_2$ and for
  different choices of~$\epsilon$ and initial states,
  we find that there are indeed ranges of these
  parameters for which the predictions of Cross are
  qualitatively correct. However, when the radius
  difference~$r_2-r_1$ becomes larger than a few roll
  diameters, a new pattern is observed consisting of
  stripe domains separated by radially-oriented grain
  boundaries. The relative stabilities of the various
  patterns are compared by evaluating their Lypunov
  functional densities.
\end{abstract}


\section{Introduction}
\label{intro}

A central question of nonequilibrium physics is pattern
formation, why only certain stationary patterns in a
nonequilibrium, homogeneous, and continuous medium are
observed for fixed parameters and boundaries. In this
paper, we investigate a particular aspect of this
question, namely how does pattern formation depend on
the shape of the lateral boundaries. We do so by
numerically integrating a two-dimensional model of a
three-dimensional convecting fluid---the
Swift-Hohenberg
equation~\cite{Swift77,Greenside84pra}---in an annular
domain as a function of the inner and outer radii of
the annulus and of the reduced Rayleigh
number~$\epsilon$ (which measures how strongly the
fluid is driven out of equilibrium).  Theory predicts
that, as parameters are varied, two kinds of patterns
may be observed, a defect-free straight roll texture or
an array of dislocations~\cite{Cross82}.  The results
we present below confirm some of these predictions and
also predict a new state of radially oriented grain
boundaries if the radial width becomes sufficiently
large. Future Rayleigh-B\'enard experiments with a
large Prandtl number fluid near onset in an annular
domain may be able to confirm these results.

Our calculations were motivated by
experiments~\cite{Heutmaker87,Croquette89I,Bodenschatz91}
and simulations~\cite{Greenside84pra} which show that a
rich variety of stationary patterns are observed just
above the onset of Rayleigh-B\'enard convection in the
form of stripes (locally periodic convection rolls)
disordered by topological defects such as dislocations,
focus singularities, and grain boundaries.
(Section~VIII of the review article by Cross and
Hohenberg~\cite{Cross93} discusses many experimental
and theoretical results for Rayleigh-B\'enard
convection.)  The occurrence of defects can be
qualitatively understood as a consequence of two
competing effects.  First, near onset the local wave
numbers of the convection rolls are constrained by the
Busse stability
balloon~\cite{Busse78,Gollub82pra,Egolf98sdc} to lie in
a narrow range about the critical wave number~$q_c$.
Second, in the absence of horizontal thermal gradients,
convection rolls are observed to evolve so as to
approach the lateral boundaries at an approximately
normal angle~\cite{Heutmaker87}. For the most commonly
used experimental geometries of cylinders or boxes,
these two constraints conflict with one another since
it is not possible for the rolls to be everywhere
normal to the sidewalls and to wiggle sinusoidally on a
single length scale. This conflict can be partially
resolved by the formation of topological defects which
allow a local change in wave vector.

An insightful analysis of this conflict between lateral
boundary conditions and local wave number was given by
Cross~\cite{Cross82}, who observed that sufficiently
close to the onset of convection, the pattern formation
was governed by a Lyapunov functional~$\mathcal{L}$
that decreases monotonically along any given orbit and
that was bounded from below.  Such a
functional~$\mathcal{L}$ implies that all initial
conditions relax asymptotically to some stationary
pattern and also provides a way to order different
stationary patterns by their relative
stability~\cite{Cross82}. In the double limit
\begin{equation}
  \label{eq:double-limit}
  \epsilon \to 0^+
  \qquad \mbox{and}\qquad
  \epsilon^{1/2} \Gamma \to \infty ,
\end{equation}
where~$\epsilon$ is proportional to the reduced
Rayleigh number $(R-R_c)/R_c$ and where~$\Gamma$ is the
aspect ratio (ratio of the largest lateral dimension to
the fluid depth), Cross showed that the Lyapunov
functional~$\mathcal{L}$ could be decomposed into a sum
of three terms, a surface term associated with the
lateral boundaries, a bulk term for the fluid far from
the boundaries, and a defect term which depends on the
particular defects present. By comparing these three
terms with one another for various geometries in the
limit Eq.~(\ref{eq:double-limit}), Cross was able to
explain rather subtle phenomena such as why a pattern
of straight parallel rolls should be observed
sufficiently close to onset in a large cylindrical or
annular domain, as opposed to a pattern of concentric
rolls with the same axisymmetry as the domain.
However, a careful comparison of Cross's calculations
with experiment~\cite{Heutmaker87} or with numerical
simulation has not been carried out and it remains
unclear to what level of accuracy the limits in
Eq.~(\ref{eq:double-limit}) need to be satisfied for
Cross's asymptotic analysis to hold. Experiments and
numerical simulations can also help to determine when
higher-order terms in the amplitude expansion destroy
the property of variational dynamics, leading to a
sustained oscillatory or chaotic time dependence.

Of particular interest to us are Cross's predictions
about possible patterns in an annular domain. For this
geometry, a fundamental frustration can arise if the
rolls are normal to both the inner and outer boundaries
since it is then not possible for the local wave number
to remain close to the critical value~$q_c$ as a
function of radius in the limit of a large radial
difference~$r_2 -r_1$. Cross predicted two
possibilities that are illustrated in Fig.~7 of his
1982 paper~\cite{Cross82}.  One was that sufficiently
close to onset, the bulk term in the Lyapunov
functional would dominate, which would favor straight
rolls that extend throughout the annular cell.  Wave
number frustration and defects are then avoided since
the rolls are not normal to the sidewalls. As a second
possibility, Cross predicted that the surface term
would be more important further above onset in which
case the rolls would orient normal to the inner and
outer boundaries.  An average wave number close to the
critical value~$q_c$ could then be arranged by
introducing an array of dislocations (an example is
shown below in Fig.~\ref{Fig-r1-eq-60-eps-eq-0.1}).  To
our knowledge, these predictions have not been tested
experimentally or numerically.

In this paper, we test these predictions by numerical
simulations of the two-dimensional time-dependent
Swift-Hohenberg equation~\cite{Swift77} in an annular
geometry as a simple opportunity to explore how lateral
boundaries affect pattern formation. The
Swift-Hohenberg equation is a widely-studied
rotationally invariant model that undergoes a
supercritical bifurcation from a zero field state to a
stripe state, corresponding to the supercritical
bifurcation of a conducting motionless fluid to
finite-amplitude three-dimensional convection rolls.
Sufficiently close to the onset of convection, the
long-wavelength slow-timescale dynamics of the
Swift-Hohenberg equation obeys the same amplitude
equation as the one obtained from the three-dimensional
Boussinesq equations that describe a convecting fluid
quantitatively~\cite{Cross93}. The linear stability
boundaries of the Swift-Hohenberg equation are also
similar to that of the Boussinesq equations near
onset~\cite{Greenside84pra,Greenside85}.  Numerical
integrations of the Swift-Hohenberg equation in
periodic, rectangular, and cylindrical
geometries~\cite{Greenside84pra,Tu92,Louie01} have
shown that its solutions are often in semi-quantitative
agreement with experiments sufficiently close to onset.
It is then often useful to explore some questions of
pattern formation first with Eq.~(\ref{Eq-2d-sh-eq})
and later make more careful studies by experiment or by
more expensive simulations of the Boussinesq equations.

Using a new time-integration code that we have
developed to integrate the Swift-Hohenberg equation
efficiently and accurately in large annular
domains~\cite{Sensoy99thesis}, we have calculated
transient solutions and their asymptotic stationary
patterns for different initial states and for different
values of the parameters~$r_1$, $r_2$, $\epsilon$. As
we describe below, our results confirm parts of Cross's
predictions but we also find that his analysis was
incomplete in that a new state occurs when the radius
difference~$r_2-r_1$ becomes sufficiently large.  This
new state consists of patches of stripes separated by
radially-oriented grain boundaries.

Using a modified version of the code, we have also
integrated a Generalized Swift-Hohenberg (GSH)
model~\cite{Manneville83:model,Greenside85} to explore
whether the wave number frustration caused by the
annular geometry may lead to a sustained dynamics.  The
GSH~model couples a second field, the vertical
vorticity potential~$\zeta(t,x,y)$, to
Eq.~(\ref{Eq-2d-sh-eq}) in such a way that the overall
dynamics is no longer variational and sustained
time-dependent solutions are
possible~\cite{Greenside88,Xi93prf,Cross96,Schmitz99}.
Our admittedly brief study suggests that a sustained
dynamics exceeding~30 or more horizontal diffusion
times ($\tau_h$) seems to occur for~$\Gamma > 15$ while
for $\Gamma < 15$, long-lived transients were observed
that decayed toward a time-independent state. However,
further studies will be needed to determine more
carefully whether any of the states are truly chaotic
(what might appear to be chaotic states in a spatially
extended system might instead be long-lived transients
whose average decay time grows rapidly with the system
size~\cite{Strain98prl}) and to characterize the
dependence of the dynamics on aspect ratio and other
parameters.
  
The rest of this paper is organized as follows.  In
Section~\ref{methods}, we summarize details of the
numerical method used to integrate the time-dependent
Swift-Hohenberg equation in an annular geometry and to
estimate the average Lyapunov
functional~$\langle\mathcal{L}\rangle$. In
Section~\ref{results}, we discuss the results of our
simulations and compare them to the asymptotic analysis
of Cross and to the Busse linear stability balloon for
the Swift-Hohenberg equation~\cite{Greenside85}.
Finally, in Section~\ref{conclusions}, we summarize our
main results. Further details are available in the
undergraduate Physics thesis of the first author, on
which this paper is based~\cite{Sensoy99thesis}.

\section{Methods}
\label{methods}
In this section, we discuss some details of how the
Swift-Hohenberg equation was integrated numerically in
a large annular domain.  The two-dimensional
time-dependent Swift-Hohenberg
equation~\cite{Swift77,Greenside84pra,Cross93}
\begin{equation}
  \label{Eq-2d-sh-eq}
  \partial_t \psi =
  \left[ \epsilon -
    \left( \nabla^2 + 1\right)^2 \right] \psi
  - \psi^3 ,
\end{equation}
determines the time evolution of a real scalar
field~$\psi(t,r,\theta)$ in a two-dimensional spatial
domain. Here $\partial_t = \partial/\partial{t}$
denotes the partial derivative with respect to time,
$\nabla^2 = \partial_x^2 + \partial_y^2 = \partial_r^2
+ r^{-1} \partial_r + r^{-2} \partial_\theta^2$ is the
two-dimensional Laplacian operator, and the
parameter~$\epsilon$ is the bifurcation parameter, such
that the zero state~$\psi=0$ becomes linearly unstable
at a critical value~$\epsilon_c(\Gamma) \ge 0$. (As
shown in Fig.~3 of Ref.~\cite{Greenside84pra}, for
rolls parallel to a wall this critical value for the
boundary conditions Eq.~(\ref{Eq-polar-bcs}) below goes
to zero as~$(2/\Gamma)^2$ in the limit~$\Gamma \to
\infty$ so~$\epsilon_c$ is tiny for~$\Gamma > 10$.)
The relation of Eq.~(\ref{Eq-2d-sh-eq}) to the
amplitude equation of the Boussinesq equations
shows~\cite{Cross82} that, near onset (i.e., in the
limit $\epsilon \to \epsilon_c^+$), the field~$\psi$ is
proportional to the temperature deviation~$T(t,x,y,z_0)
- T_{\rm linear}(z_0)$ of the temperature field~$T$
from its linear conducting profile~$T_{\rm linear}$,
evaluated on the horizontal midplane~$z=z_0$ of a
three-dimensional convection cell.  Positive and
negative values of~$\psi$ can therefore be interpreted
as warmer (rising) and cooler (descending) regions
respectively of a convecting fluid.

We solve Eq.~(\ref{Eq-2d-sh-eq}) in polar
coordinates~$(r,\theta)$ on the annular domain
\begin{equation}
  \label{Eq-annular-domain}
  r_1 \le r \le r_2, \qquad 0 \le \theta < 2\pi ,
\end{equation}
where the inner radius~$r_1$ and outer radius~$r_2$ are
prescribed parameters.  Since the scaling of parameters
that lead to Eq.~(\ref{Eq-2d-sh-eq}) normalizes the
critical wave number to have the value~$q_c=1$, the
effective fluid depth for the Swift-Hohenberg equation
is $d=(1/2)(2\pi/q_c)=\pi$ and the aspect
ratio~$\Gamma$ of the annular cell is
\begin{equation}
  \label{Eq-Gamma-defn}
  \Gamma = {r_2 - r_1 \over \pi } .
\end{equation}
For boundary conditions, we require that the
field~$\psi$ and its normal derivative~$\partial_r\psi$
vanish at each point of the inner boundary~$r=r_1$ and
at each point of the outer boundary~$r=r_2$:
\begin{equation}
  \label{Eq-polar-bcs}
  \psi(r_i,\theta) = 0, \quad \mbox{and} \quad
  (\partial_r \psi)(r_i,\theta) = 0 , \qquad
  i = 1, 2.
\end{equation}
Although these conditions are consistent with the
vanishing of the amplitude field~$A=0$ as derived to
lowest-order in an amplitude expansion
in~$\epsilon$~\cite{Cross82}, they do not correspond to
the correct boundary conditions of the amplitude~$A$
further from onset. It is then perhaps best to consider
the conditions Eq.~(\ref{Eq-polar-bcs}) as convenient
phenomenological conditions that numerous numerical
calculations~\cite{Greenside84pra,Greenside88} have
shown to yield reasonable solutions when compared with
large-aspect-ratio convection experiments near onset.

The numerical integration of Eq.~(\ref{Eq-2d-sh-eq})
with boundary conditions Eq.~(\ref{Eq-polar-bcs})
produces a discrete numerical solution~$\psi_{ijk}$ at
successive equally spaced times~$t_i = i \Delta{t}$
that approximates the unknown analytical
solution~$\psi(t,r,\theta)$ on 
mesh points $(r_j,\theta_k)$ defined by
\begin{eqnarray}
  \label{Eq-discrete-polar-mesh}
  r_j &=& r_1 + j \Delta{r}, \qquad 0 \le j \le N_r , \\
  \theta_k &=& k \Delta{\theta} , \qquad 0 \le k \le N_\theta .
\end{eqnarray}
Here~$\Delta{r} = (r_2-r_1)/N_r$ and~$\Delta{\theta}=
2\pi/(N_\theta+1)$ are the radial and angular
resolutions, and the discrete solution converges to the
unknown analytical solution at the mesh points in the
limit of vanishing space-time resolution:
\begin{equation}
  \label{Eq-psi-ijk-defn}
   \psi_{ijk} \to \psi(t_i,r_j,\theta_k) 
   \qquad \mbox{as}
   \qquad \Delta{t}, \Delta{r}, \Delta{\theta} \to 0 .
\end{equation}
We discretized Eqs.~(\ref{Eq-2d-sh-eq})
and~(\ref{Eq-polar-bcs}) on the interior and boundary
mesh points by approximating all spatial derivatives
with second-order-accurate finite differences in the
quantity~$\psi_{ijk}$ as described by Bj{\o}rstad in
Ref.~\cite{Bjorstad84circ}. For example, to approximate
the biharmonic operator~$\nabla^4$ to second-order
accuracy at a given lattice point~$(r_j,\theta_k)$, a
linear combination of 13~lattice values involving up to
second-nearest neighbors was used.

Since an explicit time-stepping method for
Eq.~(\ref{Eq-2d-sh-eq}) would impose a severe stability
constraint of the form $\Delta{t} < C
\max(\Delta{r}^4,(\Delta{r}\Delta\theta)^4)$ where~$C$
is some constant~\cite{Strikwerda89}, we chose to
integrate Eq.~(\ref{Eq-2d-sh-eq}) with a
simple-to-implement operator-splitting method that
integrates the nonlinear term explicitly and then the
linear terms in Eq.~(\ref{Eq-2d-sh-eq}) implicitly.
The absence of spatial derivatives in the cubic
nonlinear term then implies that the largest time step
allowed by the algorithm is no longer bounded by some
power of the spatial resolution, although it remains
restricted by the overall stability of the
time-splitting algorithm and by accuracy. By making
runs with different temporal resolutions for a fixed
fine spatial resolution, we found that a time step
of~$\Delta{t}=0.5$ provided a reasonable balance
between accuracy and efficiency and we used this value
for most of the integrations described in the next
section. The code was sufficiently fast on a
workstation using a 667-MHz Alpha processor that we
could integrate tens of horizontal thermal diffusion
times (many multiples of $\Gamma^2$) for our
largest~$\Gamma$ domains in just a few days and so find
approximately time-independent asymptotic patterns to
good accuracy.

The first step of our algorithm integrates the
nonlinear piece~$N[\psi]=-\psi^3$ explicitly, using a
second-order-accurate Adams-Bashforth method:
\begin{equation}
  \label{Eq-2o-adams-bashforth}
  \psi^\ast = \psi_i + {\Delta{t} \over 2} \left(
   3 N[\psi_i] - N[\psi_{i-1}] \right) .
\end{equation}
This advances the known field field values~$\psi_i$ at
time~$t_i$ (we temporarily suppress the spatial
indices~$ij$ to simplify the notation) to intermediate
field values that we denote by~$\psi^\ast$. The field
values at time~$t_{i+1}$ are then obtained by solving
the linear pieces in Eq.~(\ref{Eq-2d-sh-eq}) implicitly
using a backward-Euler method with initial
data~$\psi^\ast$, leading to the equation
\begin{equation}
  \label{Eq-implicit-biharmonic-solve}
  \left( \nabla^4 + 2 \nabla^2 + 1 + {1 \over \Delta{t}} -
  \epsilon \right) \psi_{i+1} = {1 \over \Delta{t}} \psi^\ast ,
\end{equation}
with the boundary conditions Eq.~(\ref{Eq-polar-bcs})
imposed on~$\psi_{i+1}$. This constant-coefficient
generalized linear biharmonic problem can be solved
efficiently with a second-order-accurate fast-direct
method invented by
Bj{\o}rstad~\cite{Bjorstad83,Bjorstad84,Bjorstad84circ}
which has a nearly optimal computational complexity of
$O(N_r N_\theta \log(N_\theta))$.  We used the public
Fortran77 version of the annular Bj{\o}rstad solver
available through netlib~\cite{netlib}.

The input data for the computer code are then the
parameters~$\epsilon$, $r_1$, $r_2$, $N_r$, $N_\theta$,
$\Delta{t}$, the total integration time~$T$, and some
initial discrete field~$\psi_{0jk}$ on the mesh
Eq.~(\ref{Eq-discrete-polar-mesh}) that satisfies the
discrete form of the boundary conditions
Eq.~(\ref{Eq-polar-bcs}).  Previous
studies~\cite{Greenside84pra} have shown that a
reasonable spatial resolution requires at least six
mesh points per half roll (distance of~$\pi$)
or~$\Delta{r} \le \pi/6 \approx 0.52$.  For fixed~$r_1$
and~$r_2$, we therefore chose the spatial
resolutions~$N_r$ and~$N_\theta$ to provide or to
exceed this resolution in the radial and azimuthal
directions.  (Although the resolution is uniform in
the~$r$ and~$\theta$ coordinates, in an annular cell
geometry the field~$\psi$ is approximated more
accurately near the inner radius since the density of
angular mesh points is higher there.) Many of our runs
were tested at higher spatial and temporal resolutions
(usually by a factor of~2 or~4) and our results were
verified not to change with the higher resolution.

We used four different kinds of initial
conditions~$\psi_0(r,\theta) = \psi(t=0,r,\theta)$ to
explore the dependence of the final pattern on the
initial state,
\begin{equation}
  \label{Eq-initial-conditions}
  \begin{array}{ll}
    \psi_{0,1} = a \cos(k_\theta \theta) , &
    \psi_{0,2} = a \sin(k_r r + \phi) , \\
    \psi_{0,3} = a \sin\bigl(k_x  r \cos\theta + \phi\bigr)
    , \qquad &
    \psi_{0,4} = a \eta .
  \end{array}
\end{equation}
These describe respectively radially oriented rolls
($\psi_{0,1}$), azimuthally oriented (concentric) rolls
($\psi_{0,2}$), straight rolls ($\psi_{0,3}$), and
random noise ($\psi_{0,4}$).  (We also occasionally
added small-amplitude noise to the roll initial
conditions to break possible symmetries.)  The
parameter~$a$ denotes the initial amplitude (often
chosen to be~$0.1\sqrt{\epsilon}$), the parameters
$k_\theta$, $k_r$, and $k_x$ are specified wavenumbers,
$\phi$ is some specified phase, and the $\eta$ are
uniformly distributed random numbers in the interval
$[-a,a]$. Although the second step
Eq.~(\ref{Eq-implicit-biharmonic-solve}) of the
operating-splitting method automatically makes the
solution~$\psi(t+\Delta{t})$ satisfy the discrete
boundary conditions Eq.~(\ref{Eq-polar-bcs}), we found
that convergence to a smooth solution was enhanced if
we forced the initial states
Eq.~(\ref{Eq-initial-conditions}) to satisfy the
boundary conditions by setting the field~$\psi_0$ to
zero within a distance~$\Delta{r}=\pi$ of the inner and
outer radii.

The code then generated spatiotemporal
fields~$\psi_{ijk}$ on the polar $jk$-mesh at
successive times~$t_i = i \Delta{t}$.  We visualized
the fields with contour plots at positive and negative
contours of magnitude~$\pm (1/2) \max_{j,k}
|\psi_{ijk}|$. We also plotted three scalar time series
that proved useful when determining whether some state
had become stationary.  One time series was the local
field value at the midpoint of the annulus at
angle~$\theta=0$
\begin{equation}
  \label{Eq-time-series-field-value}
  s_1(t) = \psi\left(t, {r_1+r_2\over 2},0 \right) .
\end{equation}
The second series was a global quantity, the spatially
averaged mean-square field~$\langle \psi^2 \rangle(t)$
\begin{equation}
  \label{Eq-time-series-mean-square}
  s_2(t) = {1 \over A} \int_{r_1}^{r_2} \! \, dr \,
   \int_0^{2\pi} \!\!\! d\theta
   \,\, r \left[ \psi(t,r,\theta) \right]^2
\end{equation}
where~$A = \pi(r_2^2 - r_1^2)$ is the area of the
annular domain. Near onset, one can show that
Eq.~(\ref{Eq-time-series-mean-square}) is linearly
related to the global vertical heat transport (Nusselt
number) across the fluid layer and so a plot
of~$s_2(t)$ indicates how heat transport depends on the
spatiotemporal dynamics. As a third time series, we
recorded the instantaneous value of the spatially
averaged Lyapunov
functional~$\langle\mathcal{L}[\psi]\rangle$
\begin{equation}
  \label{Eq-time-series-Lyapunov-functional}
  s_3(t) = \langle {\mathcal L}[\psi] \rangle(t)
  = {1 \over A} {\mathcal L}[\psi],
\end{equation}
where~\cite{Cross82}:
\begin{equation}
  \label{Eq-Lyapunov-defn}
    {\mathcal L}[\psi] =
    \int_{r_1}^{r_2} \!\! dr 
    \int_0^{2\pi} \!\!\! d\theta \,\, r \left[
      - \epsilon \psi^2
      + {\epsilon \over 2} \psi^4 
      + \left\{ (\nabla^2 + 1) \psi \right\}^2
    \right] .
\end{equation}
Because the mesh is uniform with respect to the~$r$
and~$\theta$ variables, the integrals in
Eqs.~(\ref{Eq-time-series-mean-square})
and~(\ref{Eq-time-series-Lyapunov-functional}) could be
approximated with a simple two-dimensional version of
the rectangle rule for approximating one-dimensional
integrals~\cite{Kincaid96}.  The integrands were first
approximated at the mesh points
Eq.~(\ref{Eq-discrete-polar-mesh}) using second-order
accurate finite-difference approximations for the
spatial derivatives, and then the mesh values were
accumulated over all the mesh points. For many initial
conditions, we verified that the time series~$s_3(t)$
in Eq.~(\ref{Eq-time-series-Lyapunov-functional})
decreased monotonically with time provided that the
space-time resolution was sufficiently fine, as should
be the case for the analytical functional associated
with Eq.~(\ref{Eq-2d-sh-eq}) and with boundary
conditions Eq.~(\ref{Eq-polar-bcs}).

After observing many patterns and their time series, we
chose empirically to define a pattern to be stationary
if each of the two global time series
Eq.~(\ref{Eq-time-series-mean-square}) and
Eq.~(\ref{Eq-time-series-Lyapunov-functional}) had
converged to three significant digits. In fact, many of
the patterns continued to evolve extremely slowly even
after many horizontal diffusion times (since the phase
of the stripe amplitude is diffusing in a large
domain), and we saw changes in the fourth significant
digit of the quantities~$\langle\psi^2\rangle$
and~$\langle\mathcal{L}\rangle$.  However, when we
integrated by a factor of ten longer for certain
representative patterns (nearly 100~$\Gamma^2$ in some
cases), we found no significant changes in the patterns
except for tiny translations or rotations of the
overall roll structure.  Although future calculations
of stationary states may benefit by using more
sophisticated algorithm such as a Newton
method~\cite{Kincaid96} applied to the time-independent
form of Eq.~(\ref{Eq-2d-sh-eq}) or by using an annular
version of a recently developed fully-implicit
method~\cite{Louie01}, the present results should
provide a good starting point for comparing theory with
experiment.

\section{Results and Discussion}
\label{results}
\subsection{Overview}

In this section we summarize results of numerical
integrations in annular domains of the Swift-Hohenberg
equation Eq.~(\ref{Eq-2d-sh-eq}) for different choices
of the bifurcation parameter~$\epsilon$, inner and
outer radii~$r_1$ and~$r_2$, and initial conditions.
Our results confirm some predictions of
Cross~\cite{Cross82} but also show reveal new patterns
that arise when the aspect ratio
Eq.~(\ref{Eq-Gamma-defn}) becomes sufficiently large.
We also mention a few results for the Generalized
Swift-Hohenberg equation which permits time-dependent
nontransient states.

The space of parameters $\epsilon$, $r_1$, $r_2$, and
initial data is too large to explore systematically so
we let the predictions of Cross, as well as practical
computational constraints guide us in our choice of
parameters. Our interest was first to explore the
region defined by the limits
Eq.~(\ref{eq:double-limit}) for which the amplitude
equation should give an accurate description of the
dynamics. For the annular domain
Eq.~(\ref{Eq-annular-domain}),
Eq.~(\ref{eq:double-limit}) plus the assumption of
Cross that the annular aspect ratio is not too large
imply the following constraints
\begin{equation}
  \label{Eq-first-r1-r2-eps-conditions}
  r_1 \sqrt{\epsilon} \gg 1, \qquad
  (r_2 - r_1) \sqrt{\epsilon} \gg 1 , \qquad
  { r_2 - r_1 \over r_1 } \ll 1 ,
\end{equation}
and we would like to satisfy these inequalities as best
as possible. A second goal was to test the specific
prediction of Cross~\cite{Cross82} that a texture
involving roll dislocations becomes more stable (has a
lower Lyapunov functional value~$\mathcal{L}$ in
Eq.~(\ref{Eq-time-series-Lyapunov-functional})) when
the following inequality is first satisfied:
\begin{equation}
  \label{Eq-Cross-criterion}
  \epsilon^{1/4} \ge {r_2 - r_1 \over r_1 } .
\end{equation}
For fixed radii~$r_1$ and~$r_2$,
Eq.~(\ref{Eq-Cross-criterion}) can be tested by making
runs for~$\epsilon$ above and below the
ratio~$[(r_2-r_1)/r_1]^4$ for the four initial
conditions Eq.~(\ref{Eq-initial-conditions}), followed
by comparing the average value of the Lyapunov
functional~$\langle{\mathcal L}\rangle$ for the final
stationary states.

Another constraint on the choice of
parameters~$\epsilon$, $r_1$, and~$r_2$ for an annular
domain comes from the Busse stability balloon for the
Swift-Hohenberg equation (see Fig.~31 in
Ref.\cite{Greenside84pra}). For locally parallel rolls
that are normal to the inner and outer walls of an
annulus, one might expect a pattern change when the
local wavenumber~$q(r)$ as a function of radius crosses
a linear stability boundary. For example, if the local
wave number~$q$ near a fixed outer radius~$r_2$ is as
small as possible consistent with the zig-zag stable
region~$q \ge 1$, then the local wave number~$q(r_1)$
enters the Eckhaus unstable region $q \ge 1 +
\sqrt{\epsilon/12}$ when
\begin{equation}
  \label{Eq-r1-unstable-criterion}
  r_1 \le { r_2 \over 1 + \sqrt{\epsilon/12} } .
\end{equation}
For~$\epsilon=0.1$ and fixed $r_2=80$ (two parameters
used in the runs below), the local wavenumber~$q(r_1)$
becomes Eckhaus unstable when $r_1 \le 73.3$. The
corresponding aspect ratio~$\Gamma \approx 2.1$ is so
small that the critical value~$\epsilon_c$ for the
instability of the zero-field state in a finite system
exceeds~$\epsilon=0.1$, i.e., as soon as a finite
amplitude radial stripe pattern can form, it is already
Eckhaus unstable. A larger outer radius~$r_2$ would
therefore be needed to test the implications of
Eq.~(\ref{Eq-r1-unstable-criterion}).

Given the above constraints and our computational
resources, we chose to fix the outer radius at the
value~$r_2 = 80$ (about 25 half-rolls) and study
patterns with a decreasing sequence of inner radii,
from~$r_1=60$ to~$r_1=10$ in steps of~10. This
corresponds to increasing the aspect ratio~$\Gamma$
from~$6.4$ to~$22.3$ in steps of~$\Delta\Gamma = 3.2$.
Eq.~(\ref{Eq-first-r1-r2-eps-conditions}) then implies
that we can make comparisons with Cross's prediction
Eq.~(\ref{Eq-Cross-criterion}) for~$\Gamma \le 10$. For
the different choices of~$r_1$ for fixed~$r_2$, we
studied just two values of the bifurcation parameter,
$\epsilon = 0.1$ and~$\epsilon = 0.5$. Earlier
calculations of the Swift-Hohenberg equation in
rectangular domains~\cite{Greenside84pra} show that
barriers to dislocation motion appear for~$\epsilon$
larger than about~$0.5$ so these two values probe the
``near onset'' and ``not so close to onset'' regimes.

\subsection{Stationary Patterns}
\label{stationary-patterns}

We begin our discussion of results with
Fig.~\ref{Fig-r1-eq-60-eps-eq-0.1}, which shows two
asymptotic stationary patterns near onset ($\epsilon =
0.1$) obtained by integrating the Swift-Hohenberg
equation for more than 15~horizontal diffusion times
(units of $\tau_h = \Gamma^2$) in a small aspect ratio
$\Gamma \simeq 6.4$ annular domain of outer
radius~$r_2=80$.  The initial conditions for panels~(a)
and~(b) were respectively radially oriented rolls and
straight rolls ($\psi_{0,1}$ and~$\psi_{0,3}$ in
Eq.~(\ref{Eq-initial-conditions})).  (Random initial
conditions lead to the same kind of state so a pattern
with dislocations seems to be the most accessible basin
of attraction.) The asymptotic patterns in both cases
consist of approximately straight radially oriented
rolls disrupted by four to six dislocations. The states
in~(a) and~(b) have an identical average Lyapunov
functional~$\langle{\mathcal L}\rangle = 3.40\times
10^{-3}$ to three significant digits and so are
extremely close in ``energy''.  Initial conditions
consisting of concentric rolls of high symmetry and a
little noise evolve into nonlinear stationary
concentric rolls which therefore constitute another
basin of attraction, although a small one.  The average
value of the Lyapunov function $\langle{\mathcal
  L}\rangle = 0.0819$ is higher than that for any of
the patterns with dislocations so the concentric rolls
are not preferred.

The value~$\epsilon = 0.1$ in
Fig.~\ref{Fig-r1-eq-60-eps-eq-0.1} substantially
exceeds the value~$[(r_2-r_1)/r_1]^4 \approx 0.012$
predicted by the criterion
Eq.~(\ref{Eq-Cross-criterion}) with~$r_2=80$, $r_1=60$
and so Fig.~\ref{Fig-r1-eq-60-eps-eq-0.1} is consistent
with Cross's prediction that states with dislocations
are preferred in that they have a lower average
Lyapunov functional. In fact, because the
threshold~$\epsilon \approx 0.012$ defined by
Eq.~(\ref{Eq-Cross-criterion}) is smaller than the
critical value~$\epsilon_c$ for the onset of convection
in such a small aspect ratio cell, we could not find a
pattern of straight rolls for these same parameters to
compare Lyapunov functional values. For the larger
value of~$\epsilon = 0.5$, which also satisfies
Eq.~(\ref{Eq-Cross-criterion}), asymptotic patterns
similar to Fig.~\ref{Fig-r1-eq-60-eps-eq-0.1} are
observed for initial conditions consisting of radially
oriented rolls or of random values. These patterns
further from onset have an average Lyapunov functional
and average square value of $\langle {\mathcal
  L}\rangle = -0.0875$ and $\langle \psi^2 \rangle =
2.79 \times 10^{-5}$. But for an initial condition
consisting of straight rolls at the critical wave
number~$q=1$, Fig.~\ref{Fig-r1-eq-60-eps-eq-0.5} shows
that a dislocation-free pattern of approximately
straight rolls can occur, although the stripes are only
approximately straight where they are nearly parallel
to the inner and outer boundaries. This pattern has
average values of $\langle {\mathcal L}\rangle =
0.0819$ and~$\langle \psi^2 \rangle = 2.74 \times
10^{-4}$ and so is not preferred, again in accordance
with the criterion Eq.~(\ref{Eq-Cross-criterion}).

We now increase the aspect ratio from~$\Gamma=6.4$
to~12.8 by decreasing the inner radius from~$r_1=60$
to~40 for fixed~$r_2=80$.  Cross's criterion
Eq.~(\ref{Eq-Cross-criterion}) for a texture with
dislocations to be preferred is now no longer expected
to hold since the annulus is many rolls wide and more
complex textures can be expected.
Fig.~\ref{Fig-r1-eq-40-eps-eq-0.1} shows two examples
of the approximately stationary patterns observed after
long times near onset ($\epsilon = 0.1$), starting from
two different initial conditions, radially oriented
rolls in~(a) and straight rolls in~(b).  The patterns
have some similarities to
Fig.~\ref{Fig-r1-eq-60-eps-eq-0.1} in that the rolls
are largely radially oriented but they now bend at a
substantial angle as they traverse the outer to inner
boundaries and a novel defect structure is observed in
the form of four or five radially-oriented grain
boundaries. These and other runs suggest that there are
many different stationary states possible for the
specified parameter values, all differing slightly in
the position and number of the grain boundaries and
possibly with the presence of a few other defects.  For
example, in addition to four grain boundaries
Fig.~\ref{Fig-r1-eq-40-eps-eq-0.1}(b) shows two
dislocations near the top and bottom of the inner
boundary. According to the values of~$\mathcal L$,
Fig.~\ref{Fig-r1-eq-40-eps-eq-0.1}(a) is slightly
preferred over~(b) but the lack of symmetry suggests
that neither is the global minimum for
Eq.~(\ref{Eq-Lyapunov-defn}).

For this same~$\Gamma=12.8$ cell, we carried out
several runs further from onset for the value~$\epsilon
= 0.5$.  Straight roll initial states lead to a
straight-roll stationary state (not shown) similar to
Fig.~\ref{Fig-r1-eq-60-eps-eq-0.5} with average values
of $\langle{\mathcal L}\rangle = -0.176$ and~$\langle
\psi^2 \rangle = 3.26 \times 10^{-5}$.
Radially-oriented and random initial conditions lead to
patterns similar to Fig.~\ref{Fig-r1-eq-60-eps-eq-0.1}
but with many more defects (with values
$\langle{\mathcal L}\rangle = -0.176$ and
$\langle\psi^2\rangle = 3.19 \times 10^{-5}$). Although
the straight-roll state has a lower average Lyapunov
value, it is not accessible from typical initial
conditions because of the barriers to dislocation
motion.

Fig.~\ref{Fig-r1-eq-20-eps-eq-0.1} shows two stationary
patterns near onset ($\epsilon = 0.1$) in a still
larger $\Gamma = 19.2$ cell, while
Fig.~\ref{Fig-r1-eq-20-eps-eq-0.1-time-series}
demonstrates the stationarity of such patterns by
showing how the time series for the average Lyapunov
functional~$\langle {\mathcal L}\rangle$ and mean
square field~$\langle \psi^2 \rangle$ have converged to
constant values. A new feature in
Fig.~\ref{Fig-r1-eq-20-eps-eq-0.1}(a) is the appearance
of a focus singularity for the first time (near the
upper right side of the annulus). Foci are a common
feature in large cylindrical and rectangular patterns
near onset but evidently don't occur in annular domains
until the aspect ratio is sufficiently large. The
stationary state Fig.~\ref{Fig-r1-eq-20-eps-eq-0.1}(b)
arises from straight-roll initial conditions. Again
there is a new feature with this larger aspect ratio in
the form of grain boundaries next to the top and bottom
outer boundary. Such grain boundaries are a common
feature in large rectangular cells near onset (for
example, see Fig.~9 of Ref.~\cite{Greenside84pra})
since rolls parallel to a sidewall are known to be
unstable near onset to transverse rolls. The defected
and straight roll patterns in
Figs.~\ref{Fig-r1-eq-20-eps-eq-0.1-time-series}(a)
and~(b) have identical average Lyapunov functionals to
three significant digits and so are equally stable.
However, the straight roll state empirically has a tiny
basin of attraction and so is not accessible for most
initial states.

We conclude this subsection with
Fig.~\ref{Fig-r1-eq-20-eps-eq-0.5}, which shows two
representative patterns in the $\Gamma = 19.2$ cell
further from onset for~$\epsilon = 0.5$.
Fig.~\ref{Fig-r1-eq-20-eps-eq-0.5}(a) grew out of
straight roll initial conditions and is a nearly
perfect array of stripes; the instability near
sidewalls towards transverse rolls does not occur at
this value of~$\epsilon$.
Fig.~\ref{Fig-r1-eq-20-eps-eq-0.5}(b) shows a state
that grew out of small amplitude random initial
conditions. The highly disordered lamellar-like pattern
is similar to that observed for the Swift-Hohenberg
equation in a large rectangular cell for~$\epsilon =
0.5$~\cite{Greenside84pra} and is a consequence of
larger barriers to the motion of dislocations. Such
frozen disordered states have not been observed
experimentally and are likely an artifact of the
Swift-Hohenberg equation, which is most physically
relevant to Rayleigh-B\'enard convection in the
limit~$\epsilon \to 0$ and large Prandtl number.

\subsection{Dynamical States of the Generalized
  Swift-Hohenberg Equation}
\label{dynamical-states}

Convection experiments close to onset of
room-temperature pressurized gases (Prandtl
number~$\approx 1$) have revealed many different
sustained time-dependent states of which spiral defect
chaos
(SDC)~\cite{Morris93,Decker94SDC,Xi95,Morris96,Cross96,Cakmur97}
has been perhaps the most intriguing and thoroughly
studied. Several experimental features of SDC remain
poorly understood, e.g., why the state occurs in the
first place, why its statistical properties are
insensitive to the aspect ratio of a cell (above some
minimum value of~$\Gamma$), and why, to the contrary,
the critical value~$\epsilon_{\rm SDC}$ above which SDC
is first observed (for a fixed experimental cell) is
sensitive to the size and geometry of the convection
cell.

To determine whether SDC may be modified by an annular
geometry and whether the wave number frustation of an
annular cell may itself be a source of sustained
dynamics, we have carried out a few exploratory
calculations with a Generalized Swift-Hohenberg
model~\cite{Manneville83:model,Greenside85} for which
prior numerical calculations in large square periodic
domains have produced spiral-defect-chaos like
states~\cite{Xi95,Cross96,WarningAboutSDC}. We used a
model of the form
\begin{eqnarray}
  \partial_t \psi &=&
  [\epsilon - (\nabla^2 + 1)^2]\psi
  - \psi^3
  - g(\bf U \rm \cdot \bf \nabla \rm)\psi
  ,  \label{Eq-gsh-psi-eq}  \\
  \left( 
    -\eta\nabla^2 +  c^2
  \right) \nabla^2 \zeta
  &=& \hat{z} \cdot
  \left[ \nabla \left( \nabla^2\psi \right)
    \times \bf \nabla \rm \psi 
  \right] ,   \label{Eq-gsh-zeta-eq}  \\
  {\bf U} &=& \nabla \times (\zeta \hat{z})
  , \label{Eq-U-defn}
\end{eqnarray} 
with boundary conditions Eq.~(\ref{Eq-polar-bcs}) for
the field~$\psi$ and boundary conditions
\begin{equation}
  \label{Eq-zeta-bcs}
    \zeta(r_1) = \zeta(r_2) = 0 ,
    \qquad
  \left. \partial_r \zeta \right|_{r_1} 
  = \left.\partial_r\zeta\right|_{r_2} = 0 ,
\end{equation} 
for the vorticity potential field~$\zeta(t,r,\theta)$.
Because Eq.~(\ref{Eq-gsh-zeta-eq}) is linear in~$\zeta$
and involves again a constant-coefficient generalized
biharmonic operator, it was straightforward to modify
the numerical algorithm of Section~\ref{methods} to
integrate Eqs.~(\ref{Eq-gsh-psi-eq})-~(\ref{Eq-U-defn})
in time.

For all of our runs, we used the parameters of
Ref.~\cite{Xi95}
\begin{equation}
  \label{Eq-SDC-parameters}
  \epsilon=0.5, \quad \eta = 1, \quad
  g = 50, \quad c^2=2 ,
\end{equation}
for which long-lived spiral-defect-chaos-like states
were observed with periodic boundary conditions. We
then fixed the outer radius to be~$r_2=80$ and made
several runs, each starting with random initial
conditions, for values of the inner radii from~$r_1=60$
to~$r_1=10$ decreasing in steps of~$-10$. (This
corresponds to aspect ratios between~$6.4$ and~$22.3$
in steps of~$\Delta\Gamma \approx -3.2$.)  For~$\Gamma
< 13$, we found that all states eventually became
stationary (using the time series of
$\langle\psi^2\rangle$ as the criterion) with the
state~$\Gamma=12.7$ taking the long time of
~$43\,\tau_h$ to become time independent.  For~$\Gamma
> 13$, the states were still dynamic up to our longest
integration times of~$50\,\tau_h$ although there was
evidence both visually and from time series that the
states were coarsening and slowly evolving to a state
other than SDC.

A representative example of the spatiotemporal dynamics
of the GSH model
Eqs.~(\ref{Eq-gsh-psi-eq})-~(\ref{Eq-U-defn}) in a
large aspect ratio $\Gamma=22.3$ annular cell is shown
in Fig.~\ref{Fig-gsh-r1-eq-10-eps-eq-0.5}. The contour
plots show the spatial structure of the field~$\psi$ at
times $t=4$, $12$, $18$, and~$24\,\tau_h$. At earlier
times (panel~a), there are a few spirals but these
eventually annihilate leaving a pattern that is more
slowly evolving, and with nearly half the pattern
consisting of foci. Two associated time series are
shown in
Fig.~\ref{Fig-gsh-r1-eq-10-eps-eq-0.5-time-series}.
They suggest that the dynamics is more complex for
about 10 horizontal diffusion times, then becomes
slower and simpler, unlike the experimentally observed
SDC state.  The approximately periodic behavior of the
midpoint field value in
Fig.~\ref{Fig-gsh-r1-eq-10-eps-eq-0.5-time-series}b
for~$t > 15\,\tau_h$ is likely caused by a slow
translation of stripes past the observation point and
does not correspond to a true time-periodic behavior.
These time series suggest that the pattern may
eventually become time independent.

Because the approximations relating the GSH model to
the Boussinesq equations are less well justified than
those that relate the Swift-Hohneberg model near onset,
especially concerning sustained chaotic states, it
would be interesting to continue these calculations not
with further integrations of the GSH model but with
integrations of the three-dimensional Boussinesq
equations in an annular domain, and to test the various
calculations with experiments in large-aspect-ratio
annular domains.

\section{Conclusions}
\label{conclusions}
The previous sections may be summarized as follows. We
have developed and applied a new computer code to study
pattern formation formation near onset of the
Swift-Hohenberg model in annular domains of varying
aspect ratios.  As pointed out by Cross~\cite{Cross82},
an annular geometry is interesting because there is an
inherent conflict near onset between the tendency of
rolls to be normal to the sidewalls and the tendency
for the rolls to be straight in the bulk. We have
investigated Cross's predictions for annular cells of
varying aspect ratio $6.4 \le \Gamma \le 22.3$, near
onset with $\epsilon = 0.1$, and further from onset
with $\epsilon = 0.5$.

For an annular domain of rather small aspect ratio
($\Gamma < 5$), the predictions of Cross hold and
stationary states with dislocations are preferred (have
a lower Lyapunov functional density) compared to
straight rolls. As the aspect ratio~$\Gamma$ becomes
larger, new patterns are observed that are
characterized by radially oriented grain boundaries. In
nearly all cases, for a given geometry and reduced
Rayleigh number~$\epsilon$, many different stationary
states are possible, usually differing only slightly in
the value of the Lyapunov density or mean square field
(Nusselt number).  The pattern that produces the global
minimum of~$\mathcal L$ for given parameters
values~$\epsilon$, $r_1$, and~$r_2$ is not known but we
conjecture to be a highly symmetric arrangement of
defects.  Straight-roll patterns are observed only when
starting from straight-roll initial conditions. Such
patterns can have a lower Lyapunov density than the
patterns with radially oriented rolls but evidently
have such a small basin of attraction that they are not
observed starting from most initial conditions.

We also explored the possibility of sustained
time-dependent states using a Generalized
Swift-Hohenberg model, for parameters such that a
long-lived spiral defect state is observed in a large
aspect ratio periodic square~\cite{Xi95}. However our
admittedly incomplete study revealed only long-lived
complex spatiotemporal transients. Further analysis,
preferably by experiment or by integrations of the
Boussinesq equations, will be needed to determine the
effect of an annular domain on time-dependent states.

We hope that the above results will stimulate
experiments to test the specific calculations reported
here, especially in large Prandtl number fluids near
onset for which the Swift-Hohenberg equation should
provide a reasonable description.

\section*{Acknowledgements}
We would like to thank Michael Cross for helpful
discussions and NSF grant DMS-9722814 and DOE grant
DE-FT02-98ER14892 for supporting this research.

\begin{figure}[htb]   
\caption{
  Stationary numerical solutions of the Swift-Hohenberg
  equation Eq.~(\ref{Eq-2d-sh-eq}) with boundary
  conditions Eq.~(\ref{Eq-polar-bcs}) in an annular
  domain of outer radius~$r_2=80$, inner
  radius~$r_1=60$, aspect ratio~$\Gamma \simeq 6.4$,
  and bifurcation parameter~$\epsilon = 0.1$. Part~(a):
  A pattern at time~$t=17\, \tau_h$ (units of
  horizontal diffusion times~$\tau_h$) that grew out of
  the initial state~$\psi_0 = 0.3 \cos(100\theta)$.
  Part~(b): a pattern at time~$19\,\tau_h$ that grew
  out of a straight-roll initial state $\psi_0 =0.3
  \cos(r\cos(\theta))$. For both patterns,
  $\langle{\mathcal L}\rangle[\psi] = 3.40 \times
  10^{-3}$ and~$\langle \psi^2 \rangle = 4.43 \times
  10^{-6}$.  The spatiotemporal resolutions were~$N_r =
  66$, $N_\theta = 1024$, and $\Delta{t} = 0.5$.}
\label{Fig-r1-eq-60-eps-eq-0.1}
\end{figure}

\begin{figure}[htb]   
\caption{
  Stationary pattern at time~$t = 25\,\tau_h$ starting
  from straight roll initial conditions $\psi_0 = 0.3
  \cos(r\cos(\theta))$ for the same physical and
  numerical parameters as in
  Fig.~\ref{Fig-r1-eq-60-eps-eq-0.1} but with a larger
  value~$\epsilon = 0.5$. Now $\langle{\mathcal
    L}\rangle = 0.0819$ and~$\langle \psi^2 \rangle =
  2.74 \times 10^{-5}$.  }
\label{Fig-r1-eq-60-eps-eq-0.5}
\end{figure}

\begin{figure}[htb]   
\caption{
  Two stationary patterns in a larger annular domain
  for parameters~$\epsilon = 0.1$, $r_1 = 40$, $r_2 =
  80$, $\Gamma \approx 12.7$, $\Delta{t}=0.5$, $N_r =
  130$ and~$N_\theta = 1024$. {\bf (a):} An
  approximately stationary pattern obtained at time~$t
  = 19\,\tau_h$ starting from azimuthally oriented
  rolls, with values $\langle{\mathcal L}\rangle =
  -4.50 \times 10^{3}$ and $\langle \psi^2 \rangle =
  3.32 \times 10^{-6}$. {\bf (b):} An approximately
  stationary pattern obtained at time~$t=22\,\tau_h$
  starting from straight roll initial conditions, with
  $\langle{\mathcal L}\rangle = -4.45 \times 10^{-3}$
  and $\langle \psi^2 \rangle = 3.38 \times 10^{-6}$.
  Both patterns show new defects consisting of radially
  oriented grain boundaries.}
\label{Fig-r1-eq-40-eps-eq-0.1}
\end{figure}

\begin{figure}[htb]   
\caption{
  Two stationary patterns in an annular domain for
  parameters~$\epsilon = 0.1$, $r_1 = 20$, $r_2 = 80$,
  $\Gamma \approx 19.2$, $\Delta{t}=0.5$, $N_r = 258$
  and~$N_\theta = 512$. {\bf (a):} An approximately
  stationary pattern obtained at time~$t = 22\,\tau_h$
  starting from radially-oriented rolls. {\bf (b):} An
  approximately stationary pattern obtained at
  time~$t=22\,\tau_h$ starting from straight roll
  initial conditions. These and similar patterns
  obtained starting from random initial conditions all
  have identical average Lyapunov functionals and
  square values to three significant digits, with
  respectively $\langle{\mathcal L}\rangle = -4.86
  \times 10^{-3}$ and $\langle \psi^2 \rangle = 2.88
  \times 10^{-6}$.}
\label{Fig-r1-eq-20-eps-eq-0.1}
\end{figure}

\begin{figure}[htb]   
  \caption{
    Time series of the spatially averaged Lypuanov
    functional~$\langle{\mathcal L}\rangle(t)$
    (Eq.~(\ref{Eq-time-series-Lyapunov-functional}))
    and mean-square value~$\langle \psi^2 \rangle(t)$
    (Eq.~(\ref{Eq-time-series-mean-square})) starting
    from low-amplitude random initial conditions for
    the parameters of
    Fig.~\ref{Fig-r1-eq-20-eps-eq-0.1}. The states are
    accurately stationary for $t > 19\,\tau_h$,
    although it is rather interesting that the two time
    series do not become stationary at the same time.}
\label{Fig-r1-eq-20-eps-eq-0.1-time-series}
\end{figure}

\begin{figure}[htb]   
\caption{
  Two stationary patterns in an annular domain for
  parameters~$\epsilon = 0.5$, $r_1 = 20$, $r_2 = 80$,
  $\Gamma \approx 19.2$, $\Delta{t}=0.5$, $N_r = 258$
  and~$N_\theta = 512$. {\bf (a):} An approximately
  stationary pattern obtained at time~$t = 2\,\tau_h$
  starting from straight rolls, with values~${\mathcal
    L} = -0.110$ and~$\langle \psi^2 \rangle = 1.59
  \times 10^{-5}$. An essentially perfect straight roll
  state is attained. {\bf (b):} An approximately
  stationary pattern obtained at time~$t=8\,\tau_h$
  starting from small-amplitude random initial
  conditions. Although there are many defects, the
  rolls are still oriented normal to the boundaries.}
\label{Fig-r1-eq-20-eps-eq-0.5}
\end{figure}

\begin{figure}[htb]   
\caption{
  Contour plots of the field~$\psi(t,r,\theta)$ at
  times~$t=6$, $12$, $18$, and~$24\,\tau_h$ from an
  integration of the Generalized-Swift-Hohenberg
  equations
  Eqs.~(\ref{Eq-gsh-psi-eq})-~(\ref{Eq-U-defn}) with
  boundary conditions Eqs.~(\ref{Eq-polar-bcs})
  and~(\ref{Eq-zeta-bcs}) in an annular domain
  with~$r_2=80$, $r_1=10$, $\Gamma\approx 22.3$ and
  starting from small-amplitude random initial
  conditions. The parameters used were~$\epsilon =
  0.5$, $\eta=1$, $c^2=2$, $g=50$, $N_r=514$,
  $N_\theta=1024$, and~$\Delta{t}=0.125$. The state is
  time dependent out to 27~$\tau_h$, the longest time
  observed. }
\label{Fig-gsh-r1-eq-10-eps-eq-0.5}
\end{figure}

\begin{figure}[htb]   
\caption{
  Time series of {\bf (a)}, the mean square
  field~$\langle\psi^2\rangle$ and of {\bf (b)}, the
  midpoint value Eq.~(\ref{Eq-time-series-field-value})
  for the time dependent solution of the GSH equations
  shown in Fig.~\ref{Fig-gsh-r1-eq-10-eps-eq-0.5}. The
  dynamics is becoming slower and simpler over time.  }
\label{Fig-gsh-r1-eq-10-eps-eq-0.5-time-series}
\end{figure}



\newpage

\centerline{\epsfxsize=4in \epsfbox{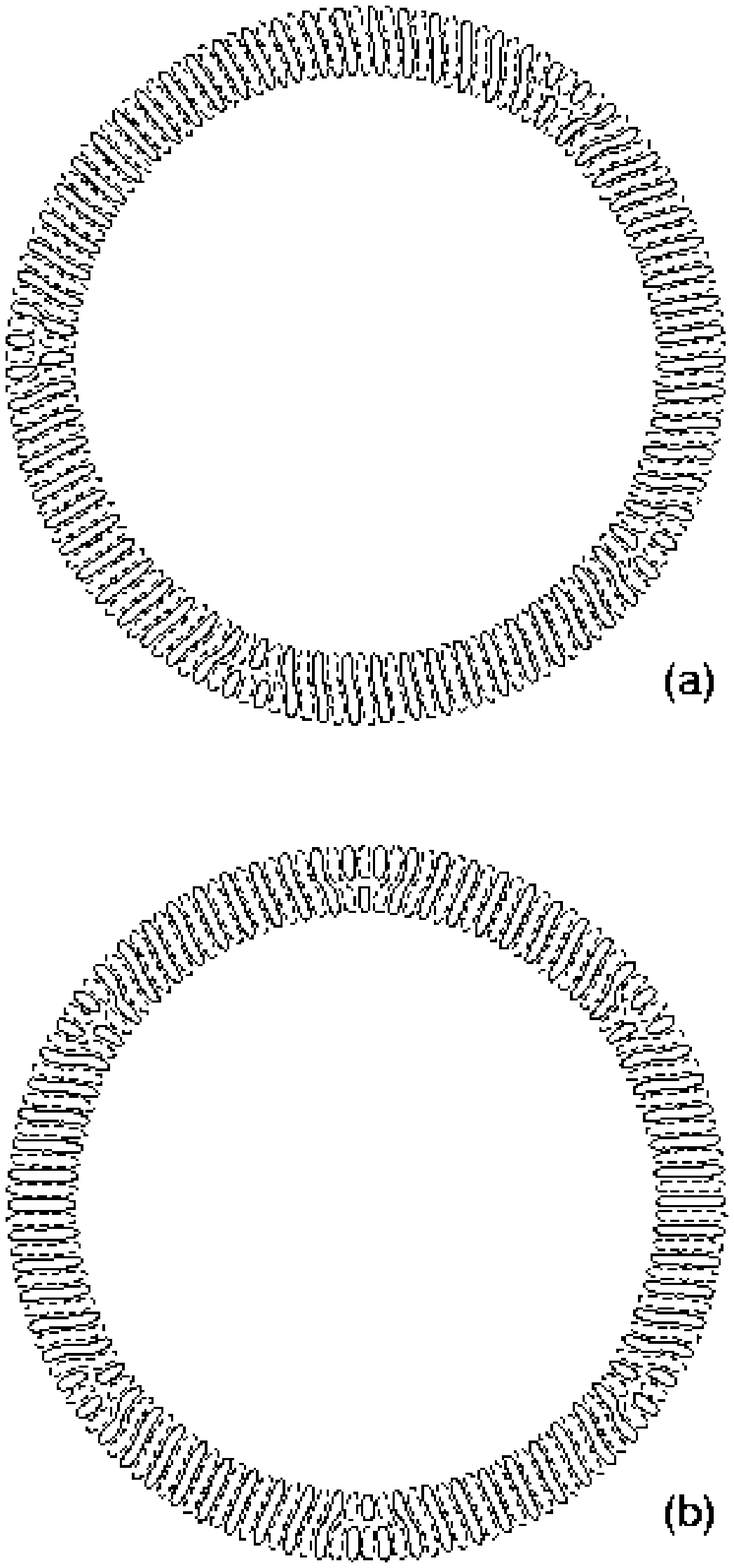}}
\begin{center}
  Figure 1.
\end{center}

\centerline{\epsfxsize=5in \epsfbox{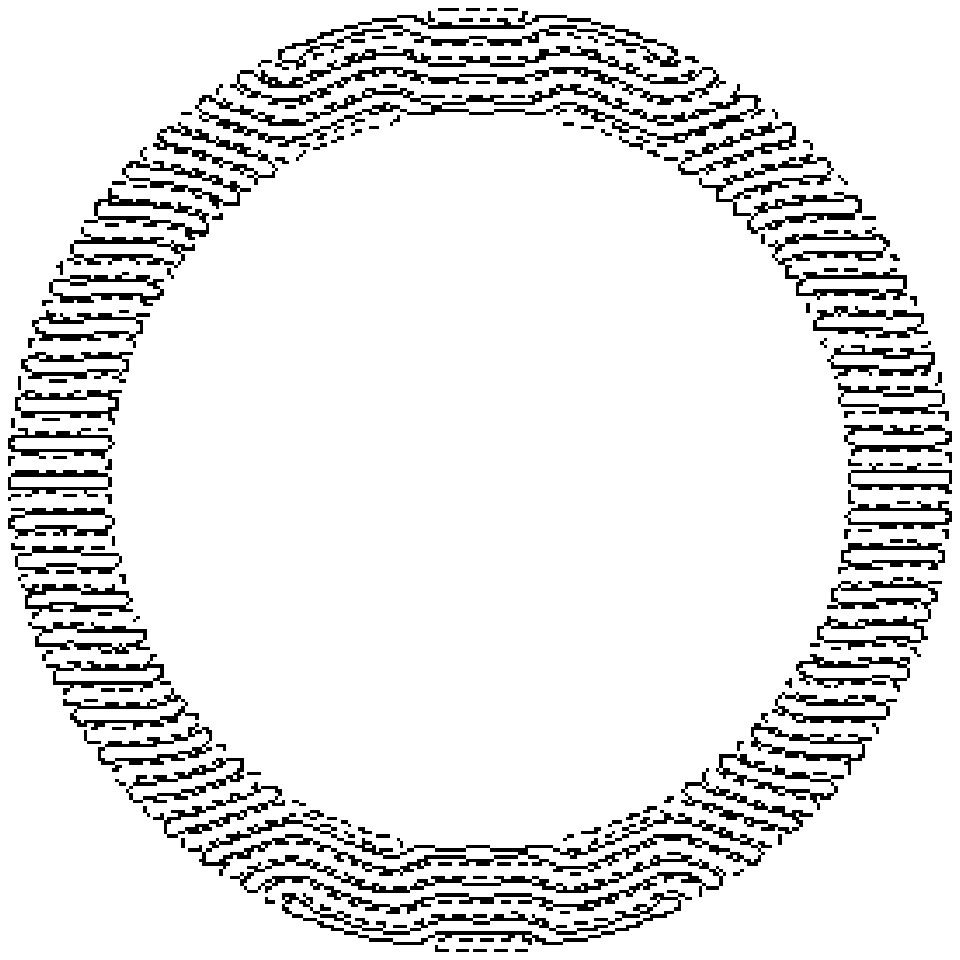}}
\begin{center}
  Figure 2.
\end{center}

\centerline{\epsfxsize=4in \epsfbox{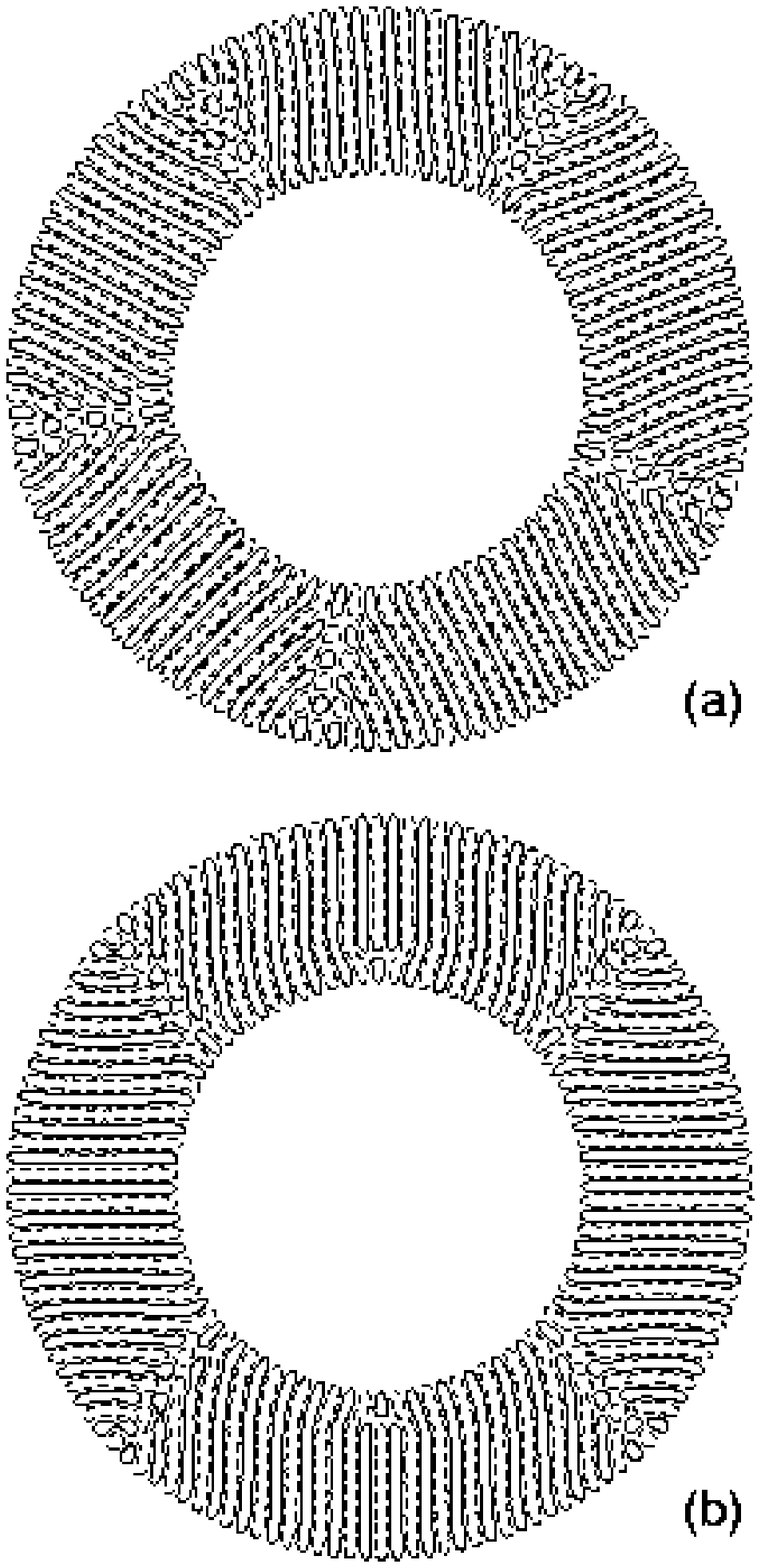}}
\begin{center}
  Figure 3.
\end{center}

\centerline{\epsfxsize=4in \epsfbox{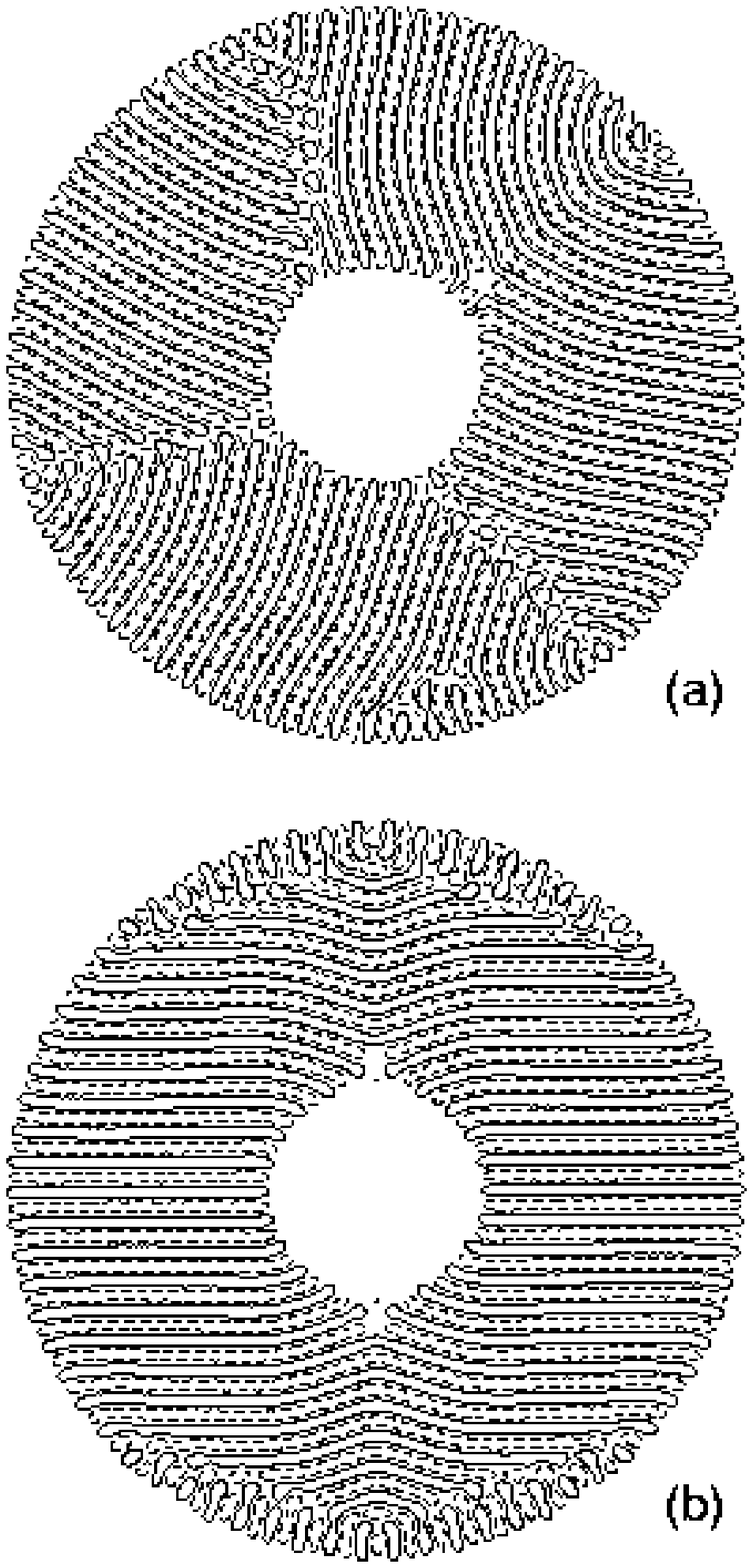}}
\begin{center}
  Figure 4.
\end{center}

\centerline{\epsfxsize=5.5in \epsfbox{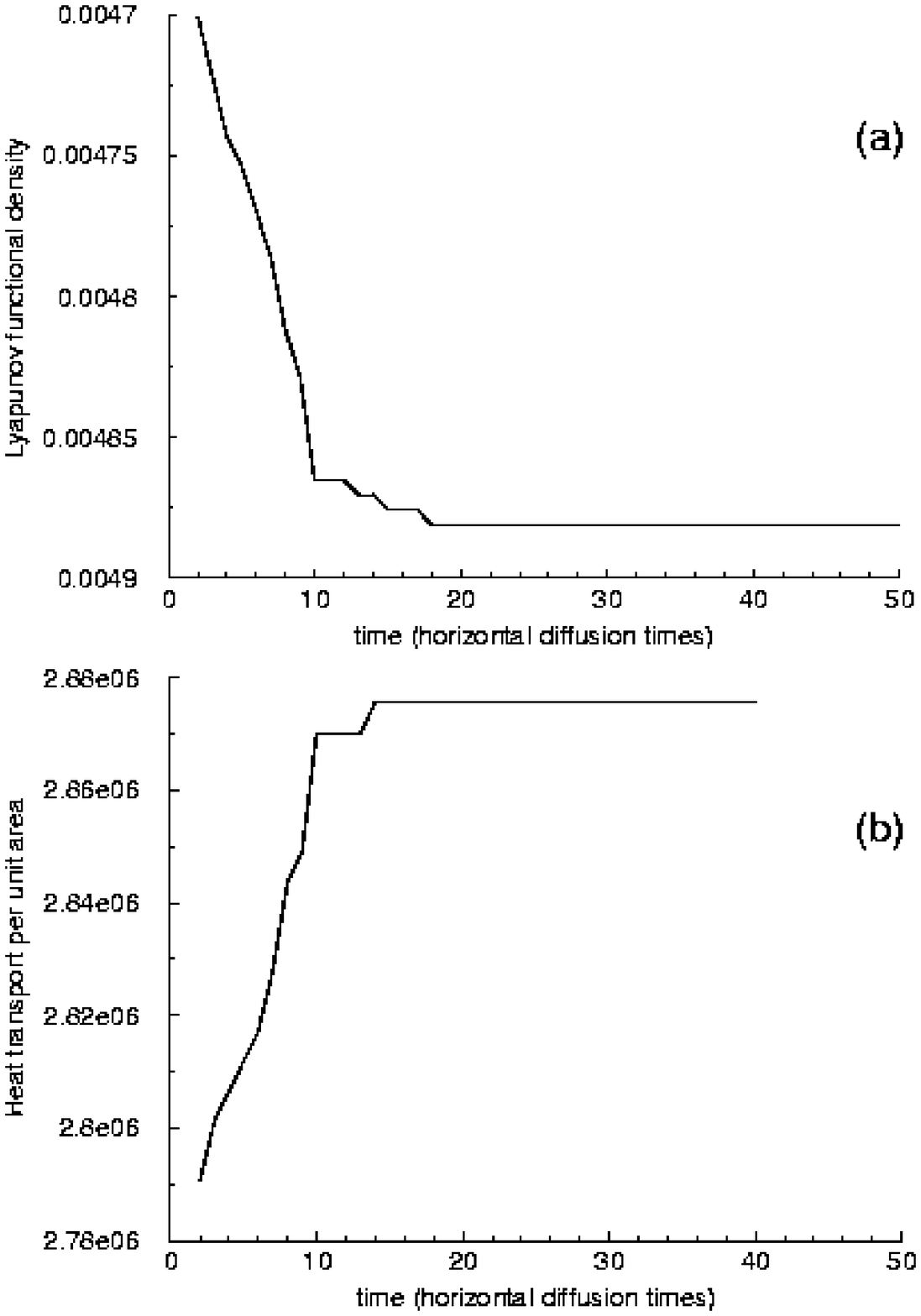}}
\begin{center}
  Figure 5.
\end{center}

\centerline{\epsfxsize=4in \epsfbox{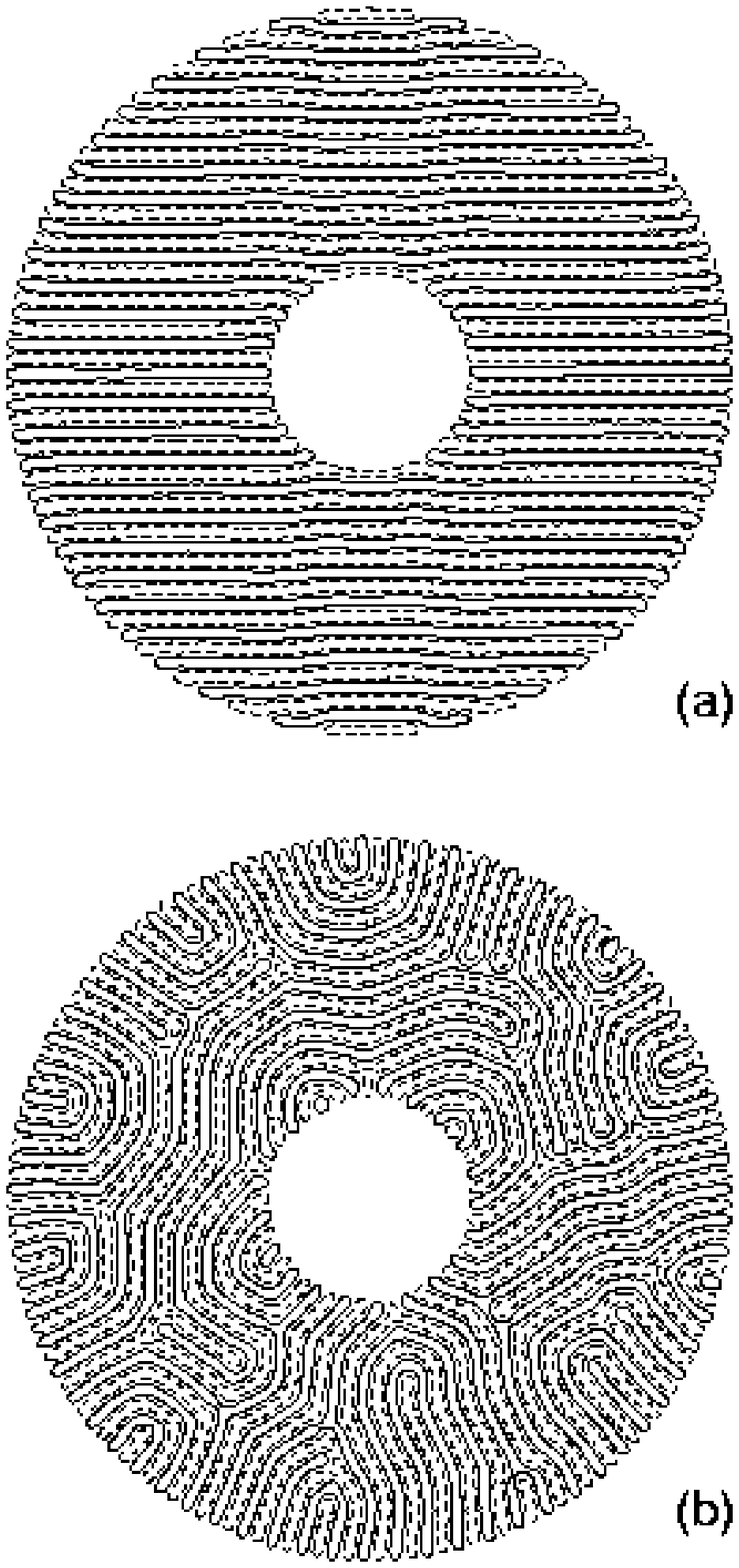}}
\begin{center}
  Figure 6
\end{center}

\centerline{\epsfxsize=6in \epsfbox{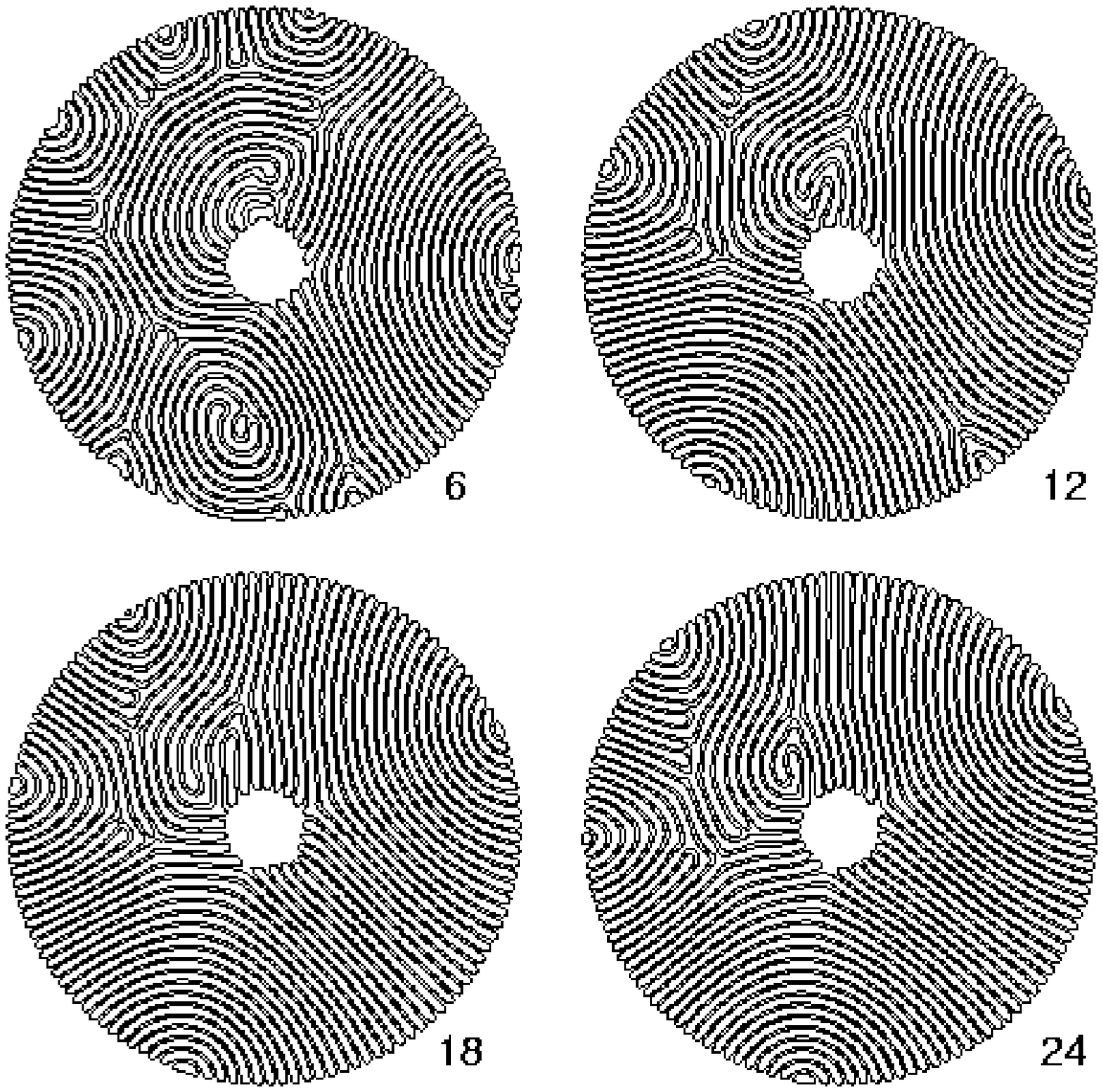}}
\begin{center}
  Figure 7.
\end{center}

\centerline{\epsfxsize=5in \epsfbox{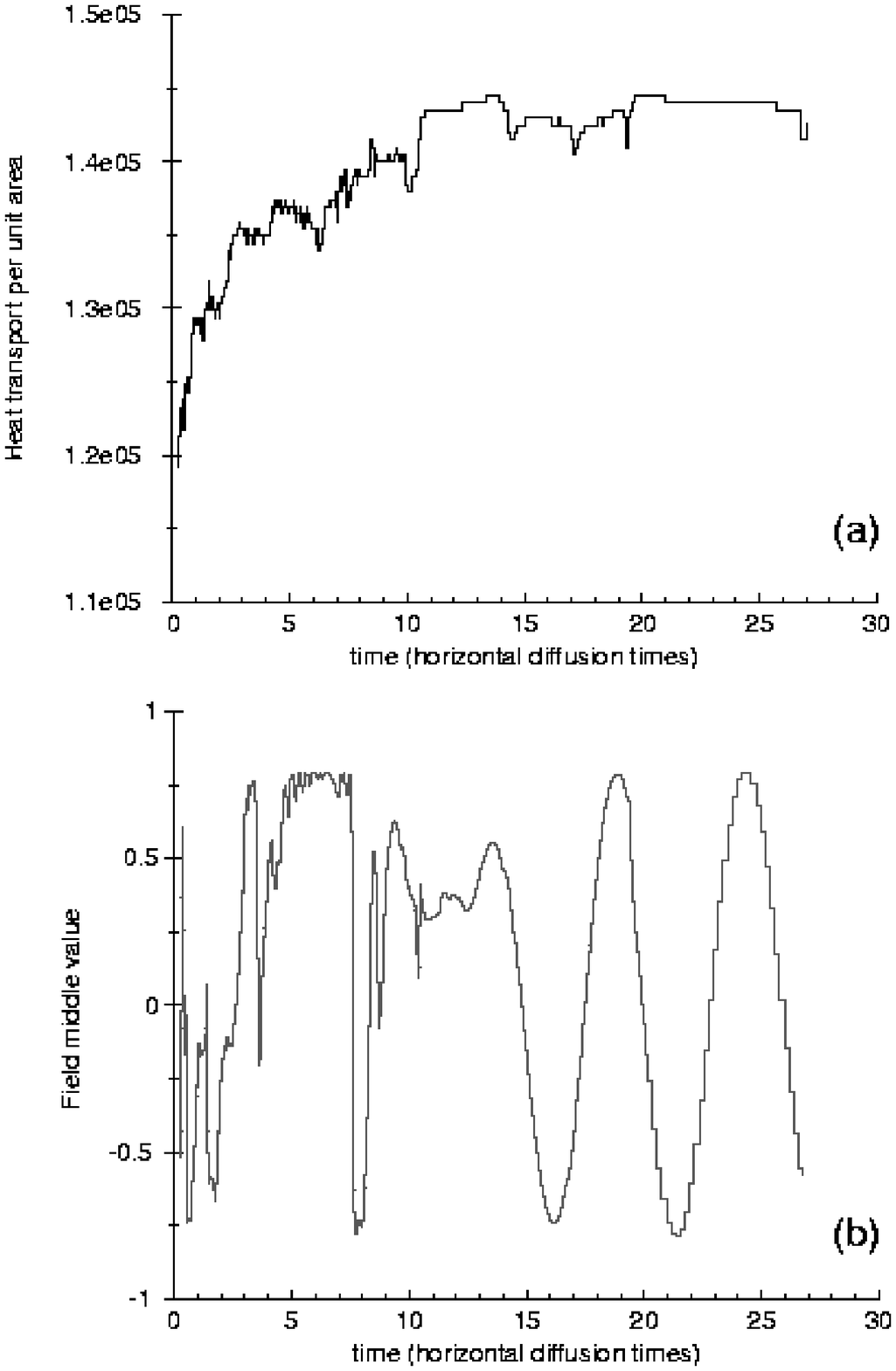}}
\begin{center}
  Figure 8.
\end{center}

\end{document}